\documentclass[%
 reprint,
superscriptaddress,
 amsmath,amssymb,
 aps,
]{revtex4-2}

\usepackage{graphicx}
\usepackage{dcolumn}
\usepackage{bm}
\usepackage{hyperref}
\usepackage{float}
\usepackage[table,xcdraw]{xcolor}
\usepackage{amsmath}

\begin{document}

\preprint{APS/123-QED}

\title{Crystal Ball: A Simple Model for Phase Transitions on a Classical Spherical Lattice}

\author{Aidan Bachmann} 
\email{abachma3@u.rochester.edu}
\affiliation{Department of Physics and Astronomy, University of Rochester}
\author{Pierre  A. Gourdain}%
\email{gourdain@pas.rochester.edu}
\affiliation{Department of Physics and Astronomy, University of Rochester}
\affiliation{Laboratory for Laser Energetics, University of Rochester}
\author{Eric G. Blackman}
\email{blackman@pas.rochester.edu}
\affiliation{Department of Physics and Astronomy, University of Rochester}
\affiliation{Laboratory for Laser Energetics, University of Rochester}

\date{\today}

\begin{abstract}
When compressed, certain lattices undergo phase transitions that may allow nuclei to gain 
significant kinetic energy. To explore the dynamics of this phenomenon, we develop a framework to study Coulomb coupled N-body systems constrained to a parametric surface, focusing specifically on the case of a sphere, as in the Thomson problem. We initialize $N$ total Boron nuclei as point particles on the surface of a sphere, allowing the particles to equilibrate via Coulomb scattering with a viscous damping term. To simulate a phase transition, we remove $N_{rm}$ particles, forcing the system to rearrange into a new equilibrium. We develop a scaling relation for the average peak kinetic energy attained by a single particle as a function of $N$ and  $N_{rm}$. For certain values of $N$, we find an order of magnitude energy gain when increasing $N_{rm}$ from 1 to 6, indicating that it may be possible to  engineer a lattice that maximizes the energy output. 

\end{abstract}

\maketitle

\section{\label{sec:intro}Introduction}
Lattice structures under compression can undergo phase transitions (for example, \cite{Hoang_2019}). During the transition, atoms of the material are rearranged into a new equilibrium, resulting in the conversion of some potential energy into kinetic energy on the time scale of the transition. To probe this phenomenon, we develop a simple, classical model to study phase transitions between equilibrium states of ion lattices. We consider ions confined to the surface of a sphere (``spherical crystals"), but the technique generalizes to other parametric surfaces. 

Equilibrium states of a single species ion lattice confined to a spherical surface are solutions to the Thomson problem. Originally posed by J.J. Thomson with respect to his ``plum pudding model", the problem is to find the configuration of N charges minimizing the Coulomb potential on the surface of a sphere. The spherical topology necessitates that these minima exhibit defects, which affect crystal properties \citep{Wales_2006}. While no analytical solutions have been found for $N > 12$, global minima candidates have been widely explored through a number of optimization techniques\citep[e.g.][]{Wales_2006,Wales_2009,Moscato_2023,Lai_2024}. However, to the authors' knowledge, phase transitions between these global minima have  yet to be studied.  Our model directly addresses this. Fusion may be another application. Since quantum tunneling enables fusion in low mass stars \cite{Balantekin_1998}, what we call
Lattice Compression Fusion (LCF) might be induced during a lattice phase transition by injecting fusion reactants into the system.  While fusion requires a fully quantum model, which is beyond the scope of this work, our classical model provides a simplifying first step toward assessing if, classically, atoms can be accelerated toward each other at high speeds. 

\begin{figure}[ht]
    \centering
    \includegraphics[width=0.9\linewidth]{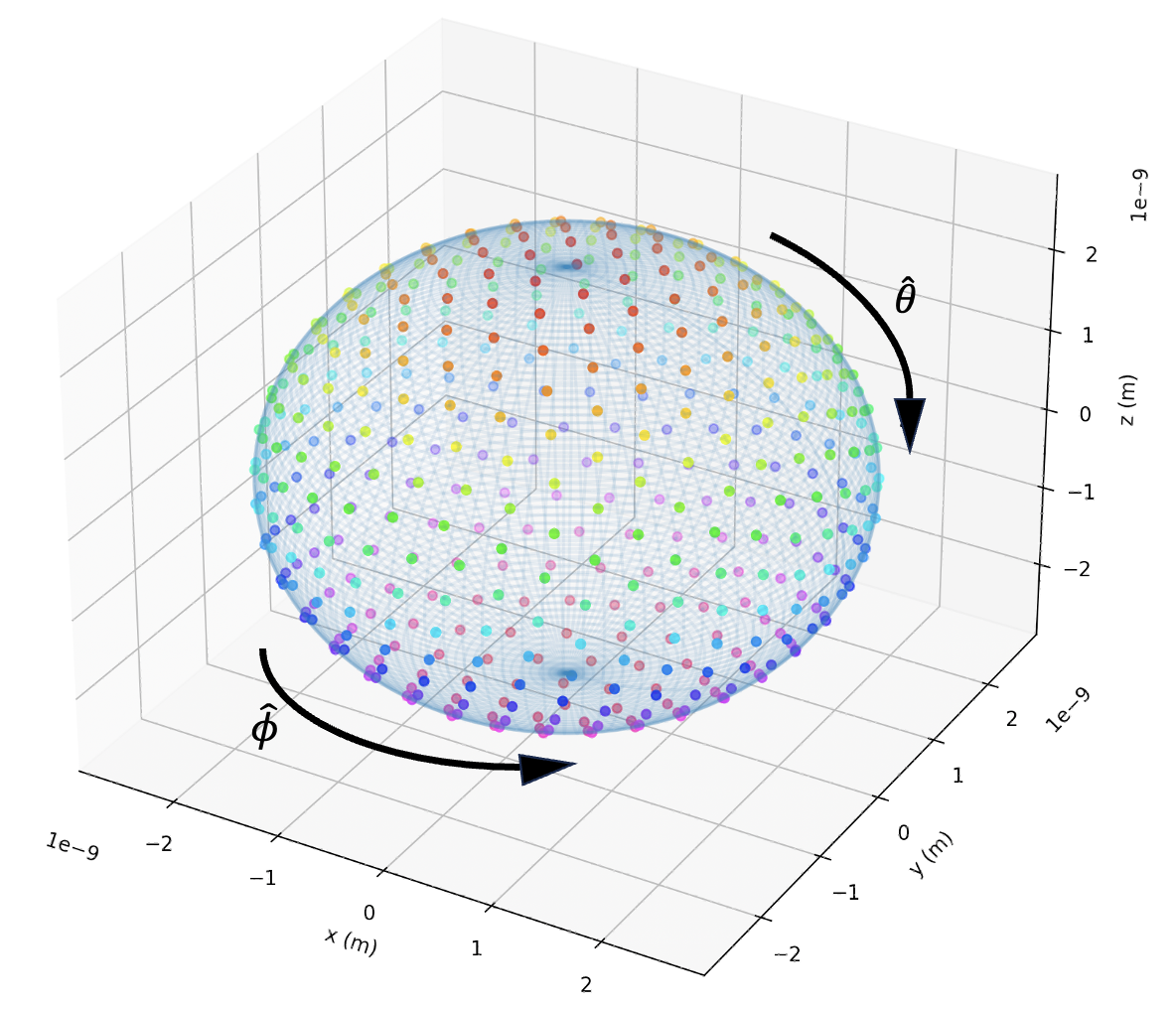}
    \caption{Example of a two dimensional spherical lattice used in the CB model. This lattice has been initialized using the method detailed below. The unit vectors $\hat{\phi}$ and $\hat{\theta}$ denote the azimuthal and polar angles, respectively.}
    \label{fig:CB}
\end{figure}

We will determine, on average, the maximum kinetic energy that one particle in the lattice can achieve over a phase transition time scale, a relevant parameter for LCF. To this end, we have developed the Crystal Ball (CB) model (Figure \ref{fig:CB}) in which $N$ particles are initialized randomly on the surface of a sphere and pushed to equilibrium through Coulomb coupling with a viscous term. To emulate the effects of a phase transition, we remove $N_{rm}$ particles from the lattice. This disrupts the periodicity, causing the system to reorganize to a new equilibrium. We then track individual particle kinetic energy and develop a scaling relation for the average peak energy $T$ as a function of $N$ and $N_{rm}$.
\section{Crystal Ball Model\label{sec:CB}}
For the CB model, we make several assumptions. In line with the Thomson problem, we ignore magnetic effects and the effects of electrons\footnote{The Thomson problem originally addresses electrons on a positive equipotential. Here, we consider a system of positive, non-relativistic point charges.}. 
We choose our point particles to be Boron nuclei with charge $Z = 5$ and mass $m \approx 11m_{p}$. LCF motivates this choice, as proton-Boron fusion is a promising candidate for aneutronic fusion\cite{Tentori_2023}. Additionally, Boron forms two-dimensional monolayers, called Borophene\cite{Kaneti_2022}\cite{Zhifen_2017}, to which our system is loosely analogous dimensionally as our lattice is an embedded two-dimensional surface. 

The Thomson problem is usually solved on a unit sphere, but we choose a non-unit radius such that the approximate lattice constant has the same order of magnitude as a Borophene lattice.  
\subsection{Initial Conditions\label{subsec:IC}}
To generate our CB lattices, we randomly place $N$ particles on the surface of the sphere and allow the system to relax. The damping term enables the system to dissipate kinetic energy and converge to an equilibrium. We chose this method both for its simplicity and because it allows the numerical integrator to be reused for phase transition simulations. For assessing convergence, we compare the potential energy of our configurations to those identified by \cite{Wales_2006} and \cite{Wales_2009}.
\subsection{Numerical Integration\label{subsec:NI}}
\begin{figure}[ht]
    \centering
    \includegraphics[width=0.7\linewidth]{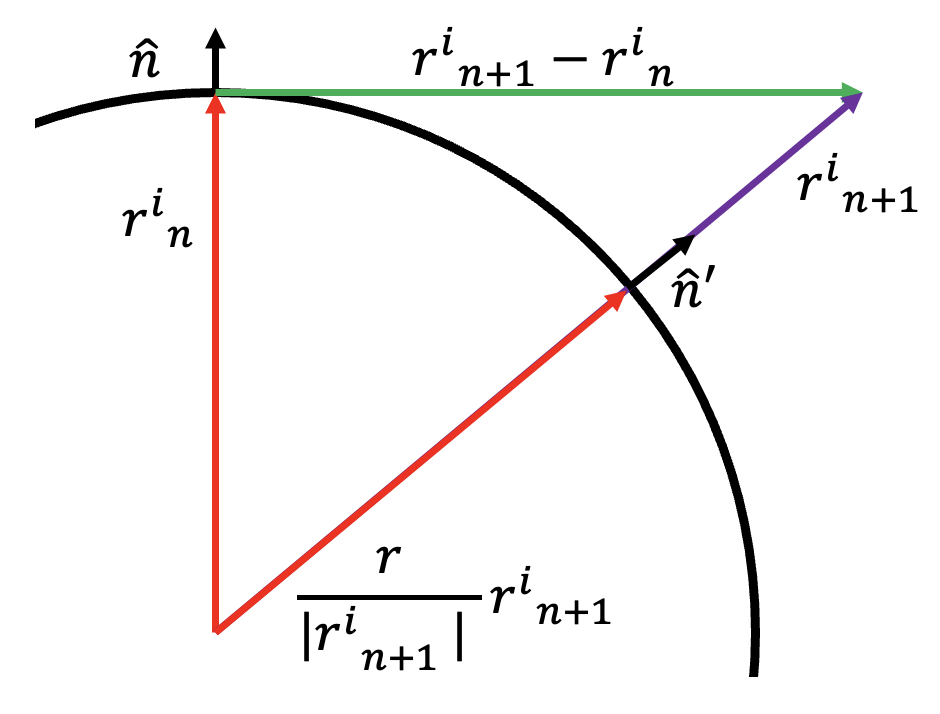}
    \caption{Updating particle position over a single time step.}
    \label{fig:algorithm}
\end{figure}

The integration scheme functions as follows. First, we compute the Coulomb force on the $ith$ particle at the $nth$ time step:
\begin{equation}
    \mathbf{F^{i}_{n,C}} = \frac{1}{4\pi\epsilon_{0}}\displaystyle\sum_{j\neq i}\frac{Z^2q^2}{|\mathbf{r^{i}_{n}}-\mathbf{r^{j}_{n}}|^3}(\mathbf{r^{i}_{n}}-\mathbf{r^{j}_{n}})\label{eq:CoulombTotal}.
\end{equation}
When generating a lattice, we impose a damping term
\begin{equation}
    \mathbf{F^{i}_{n,drag}} = -\nu \mathbf{v^{i}_{n}}^2\mathbf{\hat{v}^{i}_{n}},\label{eq:drag}
\end{equation}
where $\nu$ is the damping parameter, yielding a total force 
\begin{equation}
    \mathbf{F^{i}_{n,tot}} = \mathbf{F^{i}_{n,C}} + \mathbf{F^{i}_{n,drag}}.
\end{equation}
To constrain our particles to the surface of the sphere, we remove the component of the force normal to the sphere. The normal vector is given by
\begin{equation}
    \mathbf{\hat{n}} = \sin{\theta}\cos{\phi}\mathbf{\hat{x}} + \sin{\theta}\sin{\phi}\mathbf{\hat{y}} + \cos{\theta}\mathbf{\hat{z}},\label{eq:n}
\end{equation}
so the force tangent to the sphere is
\begin{equation}
    \mathbf{F^{i}_{n\mathbin{\!/\mkern-5mu/\!}}} = \mathbf{F^{i}_{n,tot}} - (\mathbf{F^{i}_{n,tot}}\cdot\mathbf{\hat{n}})\mathbf{\hat{n}}.\label{eq:tangent}
\end{equation}
The particle position and velocity are then updated using a first order integrator. For a time step $\Delta t$, the update is given by
\begin{equation}
    \mathbf{v^{i}_{n+1}} = \mathbf{v^{i}_{n}} + \frac{\mathbf{F^{i}_{n\mathbin{\!/\mkern-5mu/\!}}}}{m}\Delta t
\end{equation}
and
\begin{equation}
    \mathbf{r^{i}_{n+1}} = \mathbf{r^{i}_{n}} + \frac{1}{2}(\mathbf{v^{i}_{n}}+\mathbf{v^{i}_{n+1}})\Delta t + \frac{1}{2}\frac{\mathbf{F^{i}_{n\mathbin{\!/\mkern-5mu/\!}}}}{m}\Delta t^2.
\end{equation}
This update moves the particle off of the sphere surface, however the symmetry of the system allows us to re-scale the position vector's length to that of the sphere radius $R$:
\begin{equation}
    \mathbf{r^{i}_{n+1}} \rightarrow \frac{R}{|\mathbf{r^{i}_{n+1}}|}\mathbf{r^{i}_{n+1}}.\label{eq:rProj}
\end{equation}
With the particle once again lying on the sphere surface, we recompute the normal vector (it is just $\mathbf{\hat{r}^{i}_{n+1}}$) and project the velocity tangent to the sphere:
\begin{equation}
    \mathbf{v^{i}_{n+1}} \rightarrow \mathbf{v^{i}_{n+1}} - (\mathbf{v^{i}_{n+1}}\cdot\mathbf{\hat{r}^{i}_{n+1}})\mathbf{\hat{r}^{i}_{n+1}}.\label{eq:vProj}
\end{equation}
This process is repeated for some number of time steps $N_{t}$.
Additionally, an adaptive time step is used, given by
\begin{equation}
    \Delta t_{n+1} = \frac{C}{|v_{n,max}|}\frac{1}{f}\label{eq:adaptiveDT}
\end{equation}
where $C$ is the sphere circumference, $v_{n,max}$ is the maximum velocity of all particles at the $nth$ time step, and the constant $f$ is a 
tuned scaling parameter with $f\gg1$. We choose $f$ to minimize simulation time while
sufficiently conserving energy. We also impose a maximum time step to ensure numerical stability at early times. For lattice generation, $\nu$ is also 
tuned to ensure that the system is not overdamped. 
Figure \ref{fig:algorithm} shows a two dimensional visualization of this process.
A trapping potential at the sphere radius could be used instead of equations \ref{eq:rProj} and \ref{eq:vProj}, making the system Hamiltonian. However, using these projections exactly confines particles to the surface while still providing sufficient energy conservation.
\subsection{Energy Conservation}
Equation \ref{eq:vProj} reveals that our system cannot exactly conserve energy as we remove a component of the velocity at each time step, but this can be mitigated with a sufficiently small time step. Since the sphere is  locally approximately flat, the smaller the time step, the smaller the distance a particle moves off of the spherical surface each step. As a result, the energy lost to each projection decreases with smaller steps. This effect is shown in Appendix \ref{appendix:C}. As such, we will find that the energy loss in our simulations is negligible compared to the scale of the system energy.
\subsection{Lattice Initialization\label{subsec:LI}}
For our simulations, we initialized lattices for $N \in \{100,200,...,1000\}$, with a damping parameter of $\nu = 5\times10^{9}$, $f = 10^{3}$, and $R = 25.5$\AA{}. While we use Boron nuclei for our phase transition simulations, 
protons are used during initialization to allow for direct comparison to global minima of the Thomson problem as a metric for convergence. At the last time step, by assessing lattice regularity, energy of the configuration, particle velocity, and several other qualitative metrics, we find that our lattices converge for all $N$ with $N_{t} \leq 6\times10^{4}$. A full discussion is given in Appendix \ref{appendix:A}.

\section{\label{sec:sim}Scaling Relation}
Before assessing the scaling of $T(N,N_{rm})$,
we first discuss an example system that demonstrates that $T$ increases with $N_{rm}$. Note that, for any $N_{rm} > 1$, the algorithm picks a single particle at random, computes the distance to every other particle in the lattice, and removes the $N_{rm} - 1$ particles that are closest to the randomly chosen particle as well as the particle itself. Additionally, we increase $f$ to $f=10^{5}$ for all simulations.
\subsection{Example System\label{subsec:example_sys}}
For this test, we compare the results of $N_{rm} = 1$ and $N_{rm} = 7$ simulations for $N = 500$. Both systems evolve without damping for $N_{t}$ time steps. To achieve approximately the same simulation time with the adaptive time stepping scheme, we choose $N_{t} = 10^{4}$ and $N_{t} = 4\times10^{4}$ for $N_{rm} = 1$ and $N_{rm} = 7$, respectively, which sufficiently showcases the dynamics.

The results are plotted in Figure \ref{fig:ex_N_500}. Snapshots in time for both simulations can be found in Appendix \ref{appendix:B}.
\begin{figure}[ht]
    \centering
    \includegraphics[width=\linewidth]{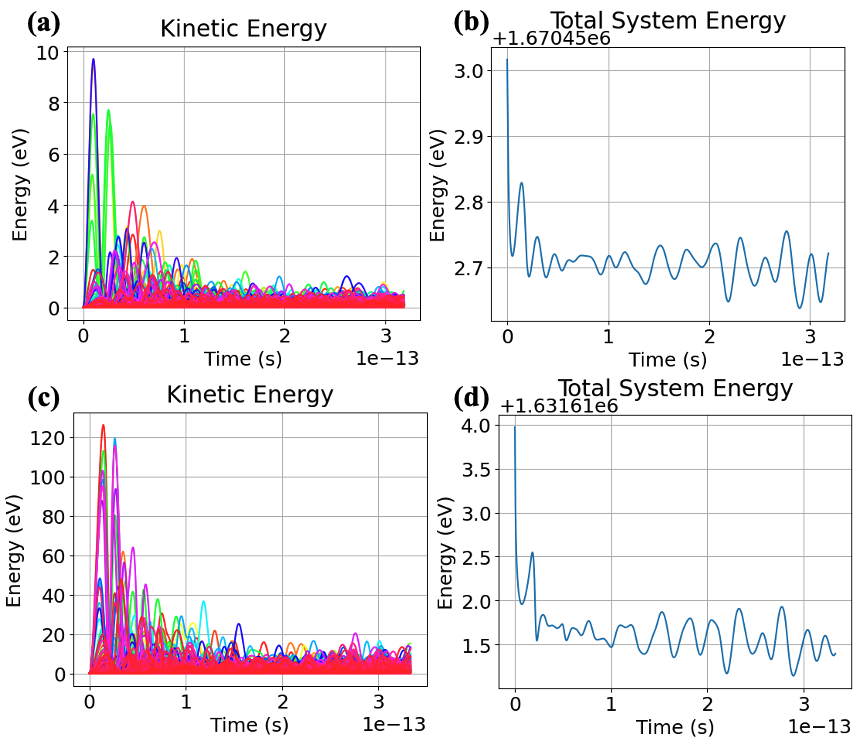}
    \caption{$N = 500$ example system. (a) $N_{rm} = 1$ kinetic energy. (b) $N_{rm} = 1$ total energy. (c) $N_{rm} = 7$ kinetic energy. (d) $N_{rm} = 7$ total energy. Color indicates particle index. We aim to understand the average scaling of the peak values in (a) and (c).}
    \label{fig:ex_N_500}
\end{figure}
We plot the kinetic and total energy time series for $N_{rm} = 1$ in (a) and (b), respectively. The analogous plots for $N_{rm} = 7$ are shown in (c) and (d). Color indicates the particle index. Both cases exhibit the same qualitative features. We see the kinetic energy peak at early times before settling down to a low energy background corresponding to oscillations around a new equilibrium. Additionally, the total system energy does decrease, however this decrease is negligible for two reasons. Firstly, the energy decrease is, in both cases, on the order of 2.5eV or less, which is negligible compared to the system energy of $\sim 10^{6}$eV. Secondly, this energy is lost by the entire system of $N-N_{rm}$ particles, indicating that on average, each particle loses at most $\sim0.002$eV, four orders of magnitude smaller than our lowest peak energy. As a result, we conclude that energy is sufficiently conserved.

In these plots, the peak kinetic energy that any one particle attains is the relevant feature for our scaling. For the case of $N_{rm} = 1$, Figure \ref{fig:ex_N_500}a indicates that this value is $\sim10$eV. Increasing to $N_{rm} = 7$, however, increases this peak energy by an order of magnitude to $\sim120$eV, which matches our expectations. We will now assess how this peak energy scales on average. We will derive our model, describe how we generate our statistics, then discuss the fit. To motivate our derivation, we will first describe a simplified 1D system.

\subsection{1D Model}
\begin{figure}[ht]
    \centering
    \includegraphics[width=\linewidth]{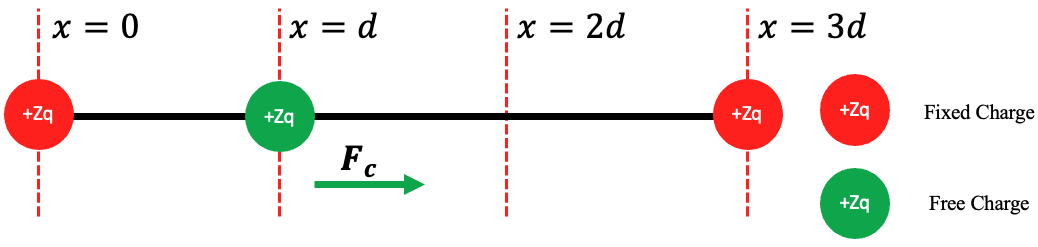}
    \caption{Initial conditions for 1D problem.}
    \label{fig:1D_prob}
\end{figure}

Consider a system of 4 equal charges $Zq$ placed with uniform spacing $d$ along a line. We fix the two charges at the end points ($x = 0$ and $x = 3d$) and remove one of the inner charges ($x = 2d$), leaving the remaining charge free (See Figure \ref{fig:1D_prob}).
Evolving the system, we can compute the free particle's peak energy by integrating the force along the path from the initial position to the fixed charges' null field point. The force is given by
\begin{equation}
        \bold{F} = \frac{Z^2q^2}{4\pi\epsilon_{0}}\bigg(\frac{1}{x^2} - \frac{1}{(x - 3d)^2}\bigg)\bold{\hat{x}}.\label{eq:1DF}
\end{equation}
By symmetry, we know $\bold{F} = 0$ at $x = \frac{3d}{2}$, so the peak kinetic energy $T$ is the work:
\begin{equation}
     T = \int_{d}^{\frac{3d}{2}}\bold{F}\cdot\bold{dx} = \frac{Z^2q^2}{4\pi\epsilon_{0}}\int_{d}^{\frac{3d}{2}}\bigg(\frac{1}{x^2} - \frac{1}{(x - 3d)^2}\bigg)dx\label{eq:1D_work_integral},
\end{equation}
which yields
\begin{equation}
    T = \frac{1}{6}\frac{1}{4\pi\epsilon_{0}}\frac{Z^2q^2}{d}\label{eq:1DT}.
\end{equation}
Equation \ref{eq:1DT} is the maximum potential energy of a particle scattering off of a like charge with a distance of closest approach of $6d$. For CB, removing $N_{rm}$ charges takes the configuration out of equilibrium, analogously to Figure \ref{fig:1D_prob}, which causes the system to do some amount of work to rearrange itself. Some fraction of this total work goes to accelerating a single charge to the peak kinetic energies seen in Figure \ref{fig:ex_N_500}. So, in analogy to \ref{eq:1D_work_integral}, we want to write the CB scaling in the form
\begin{equation}
    T = cW_{sys},\label{eq:T_wsys}
\end{equation}
where $W_{sys}$ is the total work done by the system and $c < 1$ is the work fraction giving the peak kinetic energy. With equation \ref{eq:T_wsys}, we have recast the problem so that, instead of solving for $T$ directly, we can estimate the total amount of work done by the system to compute $T$. For the 1D problem, we can compute $W_{sys}$ exactly (equation \ref{eq:1DT}) with $T$ following immediately for $c = 1$. This is much more complicated for CB, however we can use several approximations to derive $W_{sys}$ for this case.

\subsection{Scaling Relation Derivation\label{subsec:scaling}}
Assuming the system is sufficiently localized, we can estimate $W_{sys}$ by considering only the dynamics of charges directly adjacent to the ``hole" left by removing $N_{rm}$ charges. We will refer to this charge as $q_{NN}$, the total charge of $NN$ nearest neighbors. Additionally, instead of estimating the effect of all the other charges in the lattice pushing on $q_{NN}$, it is convenient to imagine that the hole has acquired some negative charge which pulls the nearest neighbors inwards. We will refer to this as $q_{hole}$. With this view, $q_{NN}$ and $q_{hole}$ are functions of $N_{rm}$. The force between $q_{NN}$ and $q_{hole}$ is determined approximately by their initial separation $d$, the average lattice constant, which is a function of $N$. Assuming that the force is approximately constant over $d$, which is also the length scale over which charges accelerate, we can write
\begin{equation}
    W_{sys} = \frac{1}{4\pi\epsilon_{0}}\frac{q_{hole}(N_{rm})q_{NN}(N_{rm})}{d(N)}.\label{eq:Wsys}
\end{equation}
All particles in our lattice have charge $Zq$, so computing $q_{hole}$ and $q_{NN}$ reduces to a counting problem. In the case of $q_{hole}$, we are removing $N_{rm}$ particles of charge $Zq$, so $q_{hole}$ is simply
\begin{equation}
    q_{hole}(N_{rm}) = N_{rm}Zq. \label{eq:qhole}
\end{equation}
We can estimate $q_{NN}$ by noticing that when we remove one particle, the resulting hole has $NN = 6$ on average. Starting with $N_{rm} = 1$, iteratively removing all particles directly adjacent to the hole and counting the average $NN$, we find that $NN = 6k$ where $k$ is the iteration. Parameterizing $N_{rm}$ as a function of $k$ gives
\begin{equation*}
    N_{rm}(k) = 3k^{2} - 3k + 1\label{eq:Nrm_k}.
\end{equation*}
Since $k = \frac{NN}{6}$, we can plug this into the equation above and solve the quadratic for $NN(N_{rm})$. This yields
\begin{equation}
    NN(N_{rm}) = 3 + \sqrt{3(4N_{rm}-1)}\label{eq:NN_Nrm},
\end{equation}
where we retain only the physically meaningful, positive root. This expression allows us to smoothly interpolate $NN(N_{rm})$ between the discrete values given by $NN = 6k$. With equation \ref{eq:NN_Nrm}, we can write
\begin{equation}
    q_{NN}(N_{rm}) = NN(N_{rm})Zq = \Big(3 + \sqrt{3(4N_{rm}-1)}\Big)Zq.\label{eq:qNN}
\end{equation}
It is important to note that equation \ref{eq:qNN} is only valid when $N_{rm}\leq\frac{N}{2}$. With the points being approximately uniformly distributed over the sphere, in the neighborhood of $N_{rm} = \frac{N}{2}$, the number of nearest neighbors begins to decrease as $N_{rm}$ increases. Thus, our scaling relation is no longer valid in this regime, so we only explore the limit of $N_{rm} \ll N$.

\begin{figure}[ht]
    \centering
    \includegraphics[width=\linewidth]{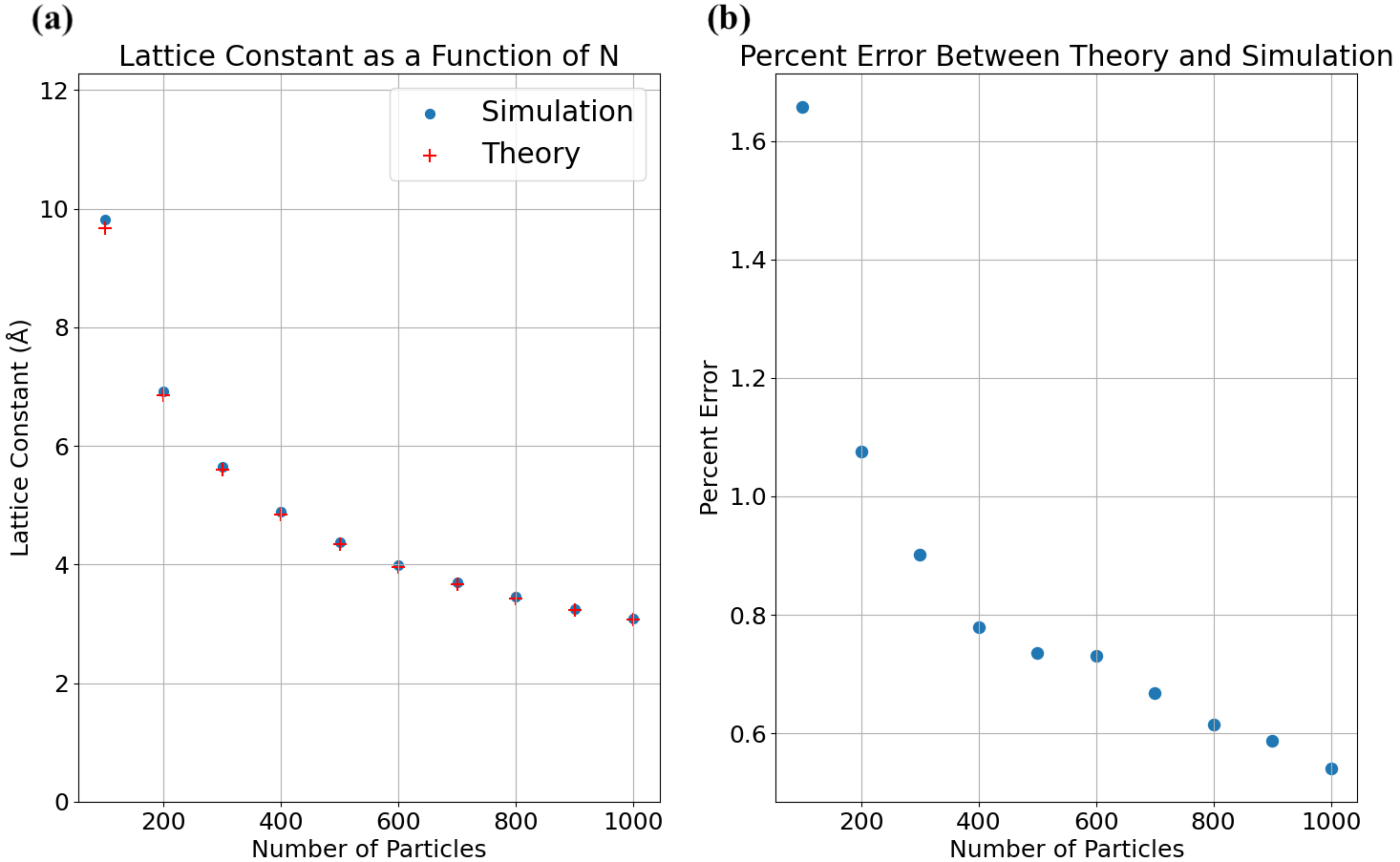}
    \caption{(a) Theoretical and simulation values of the average lattice constant (b) Percent difference between theory and simulation. The difference is nearly monotonically decreasing. This is due, in part, to overdamping, but is primarily a result of averaging over imperfections.}
    \label{fig:d_scaling}
\end{figure}

To compute the average particle spacing, we partition the surface area of the sphere into spherical triangles that are approximately equilateral. Ignoring imperfections, we can estimate the area of the sphere by summing up the area of these equilateral triangles, which we can then use to solve for $d$. We know from the Euler characteristic that a convex polyhedron with $N$ vertices has $2N - 4$ faces. Then, if a spherical triangle has area $a$, we can write
\begin{equation}
    (2N - 4)a \approx 4\pi R^2\label{eq:tessellation}
\end{equation}
for a sphere of radius $R$. Moving forward, we will write all equations as equalities even though this is not strictly correct. From Girard's theorem, we know that a spherical triangle has area
\begin{equation}
    a = R^2(\alpha + \beta + \gamma - \pi),\label{eq:girardTheorem1}
\end{equation}
where $\alpha$, $\beta$, and $\gamma$ are the interior angles of the spherical triangle. In an equilateral spherical triangle, $\alpha = \beta = \gamma$, so this becomes
\begin{equation}
    a = R^2(3\alpha - \pi).\label{eq:girardTheorem2}
\end{equation}
Plugging equation \ref{eq:girardTheorem2} into equation \ref{eq:tessellation} and solving for $\alpha$ yields
\begin{equation}
    \alpha = \frac{N}{N-2}\frac{\pi}{3}.\label{eq:alpha}
\end{equation}
Here, the factor $\frac{N}{N-2}$ acts as a correction to the interior angle of a flat triangle. Then, from the law of spherical cosines, we have that
\begin{equation}
    \cos{\frac{s}{R}} = \cos^2{\frac{s}{R}} + \sin^2{\frac{s}{R}}\cos{\alpha},\label{eq:sphericalCosines}
\end{equation}
where $s$ is the arc length of one side of the triangle. Rearranging equation \ref{eq:sphericalCosines} and solving for $s$ yields
\begin{equation}
    s = R\arccos{\bigg(\frac{\cos{\alpha}}{1-\cos{\alpha}}\bigg)}.\label{eq:s_alpha}
\end{equation}
Plugging in for $\alpha$ gives us
\begin{equation}
    s = R\arccos{\bigg(\frac{\cos{\frac{N\pi}{3(N-2)}}}{1-\cos{\frac{N\pi}{3(N-2)}}}\bigg)}, \label{eq:s_N}
\end{equation}
the average arc length distance between adjacent particles. However, since the electrostatic interaction is a central force, we need to compute the linear distance $d$. Some trigonometry yields
\begin{equation}
    d(N) = R\sqrt{\frac{2(1-2\cos{\frac{N\pi}{3(N-2)}})}{1-\cos{\frac{N\pi}{3(N-2)}}}} \label{eq:d_N}.
\end{equation}
Figure \ref{fig:d_scaling} shows how equation \ref{eq:d_N} compares to simulation values computed from the average of the distance between each particle's 6 nearest neighbors.
From the figure, we see that the theoretical value of $d$ shows good agreement with our simulations, with error not exceeding 1.7\%. We also see that the error is nearly monotonically decreasing as a function of $N$. This results partly from overdamping, but is primarily due to impurities in the lattice structure which our calculation effectively averages over. However, as $N$ increases, the number of these imperfections becomes an increasingly small fraction of the total structure, meaning that our approximation becomes more accurate for increasing $N$. This is shown in Table 1 of \cite{Wales_2009}. For our purposes, the error from this effect is sufficiently small for us to proceed with equation \ref{eq:d_N}. Plugging our expressions for $d_{N}$, $q_{NN}$, and $q_{hole}$ into equation \ref{eq:Wsys} and adding our constant $c$, we find
\begin{align}
\begin{split}
    T(N,N_{rm}) & = cW_{sys}(N,N_{rm}) \\ & = c\frac{Z^2q^2}{4\pi\epsilon_{0}R}\sqrt{\frac{1-\cos{\frac{N\pi}{3(N-2)}}}{2(1-2\cos{\frac{N\pi}{3(N-2)}})}}\times \\ & N_{rm}\Big(3 + \sqrt{3(4N_{rm}-1)}\Big).\label{eq:scaling}
\end{split}
\end{align}
We can repackage equation \ref{eq:scaling} into a form that better aligns with equation \ref{eq:1DT}:
\begin{align}
\begin{split}
    T(N,N_{rm}) &= \frac{1}{4\pi\epsilon_{0}}\frac{Z^2q^2}{\frac{d(N)}{cN_{rm}\Big(3 + \sqrt{3(4N_{rm}-1)}\Big)}} \\ &= \frac{1}{4\pi\epsilon_{0}}\frac{Z^2q^2}{L}\label{eq:scaling_binary}
\end{split}
\end{align}
where
\begin{equation}
    L = \frac{1}{cN_{rm}\Big(3 + \sqrt{3(4N_{rm}-1)}\Big)}d(N).\label{eq:L}
\end{equation}
Written in this way, our scaling relation takes on the form of the maximum potential energy of a binary interaction with distance of closest approach $L$. The effects of $q_{hole}$, $q_{NN}$, and $c$ are rolled into the characteristic length scale $L$ as a correction to the lattice constant. Note that the pre-factor in equation \ref{eq:L} is monotonically decreasing, with $L = d(N)$ for the non-physical value $N_{rm} \sim 4.76$.

\section{Conclusions and Outlook\label{sec:conclusions}}
For each initialized lattice (Section \ref{subsec:LI}), we tested $N_{rm} \in \{1,2,...,10\}$. For each $(N,N_{rm})$ pair, we performed 24 Monte Carlo (MC) simulations to generate our statistics. With $c$ as a free parameter, we fit equation \ref{eq:scaling} to the average of these simulations using a $\chi^2$ minimizer, with the standard deviation as the error.

\begin{figure}[ht]
\centering
\begin{minipage}{.45\textwidth}
  \centering
  \includegraphics[width=1\linewidth]{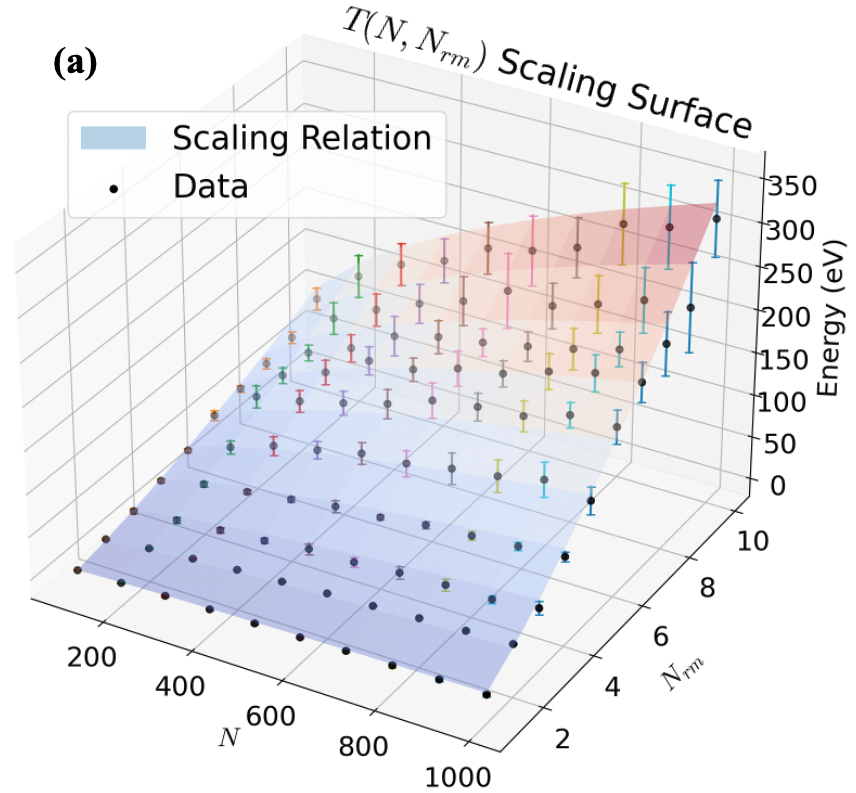}
\end{minipage}%
\hspace{0.05\textwidth}
\begin{minipage}{.45\textwidth}
  \centering
  \includegraphics[width=1\linewidth]{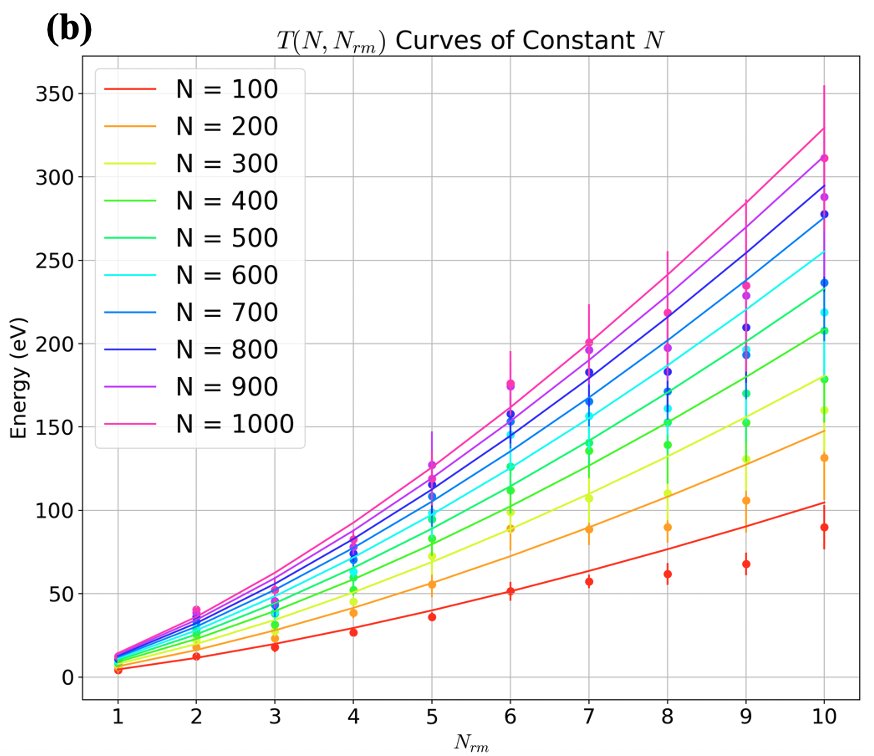}
\end{minipage}%
\caption{(a) Fitted $T(N,N_{rm})$ surface. (b) Curves of constant $N$ of $T(N,N_{rm})$. For both plots, points are average peak values of all MC simulations with error bars being the standard deviation. Fits are indicated with solid lines. For a model with $N\times N_{rm} - 1$ degrees of freedom, the fit gives a reduced $\chi^2_{fit} = 2.365$.}
\label{fig:E_scaling}
\end{figure}

Both the fitted surface $T(N,N_{rm})$ and curves of constant $N$ are shown in Figure \ref{fig:E_scaling}.
Qualitatively, the fits give a good order of magnitude scaling. Additionally, for our model with $N\times N_{rm} - 1$ degrees of freedom, the fit has a reduced $\chi_{fit}^2 = 2.365$, indicating a reasonable fit with $c = 0.0203$. Notably, our scaling does not explain the inflection point near $N_{rm} = 6$, which could be a result of the fact that $NN = 6$ on average. Away from $N_{rm} = 6$, the fit grows as the model predicts.  

This scaling suggests that it may be possible to engineer lattice structures to maximize kinetic energy output, an important first step in exploring LCF. However, to properly probe LCF, a quantum model using tools such as R-Matrix theory\cite{Burke_2011} or Density Functional Theory is necessary. Beyond LCF, the CB model itself has not been fully characterized. Particularly, we have not described energy transport in the lattice, which our results indicate may be a combination of wavelike and diffusive behavior. Properly understanding this mechanism is essential for describing phase transitions between equilibrium states of the Thomson problem. Similarly, we require a more sophisticated model to capture the nuances of the energy scaling. Additionally, we have not extended the model to other geometries, which simply requires a parameterization of the surface normal vector to implement. Each of these aspects are potential avenues for future work. 

\begin{acknowledgments}
The authors would like to thank James Young and Rayleigh Parker for their comments, suggestions, and edits. This work was funded by MACH NNSA Center Funding DE-NA0004148, NSF PHY-1943939, and NSF PHY-2020249.
\end{acknowledgments}

\appendix

\section{Crystal Ball Convergence and the Thomson Problem\label{appendix:A}}
\begin{figure}[ht]
\centering
\includegraphics[width=1\linewidth]{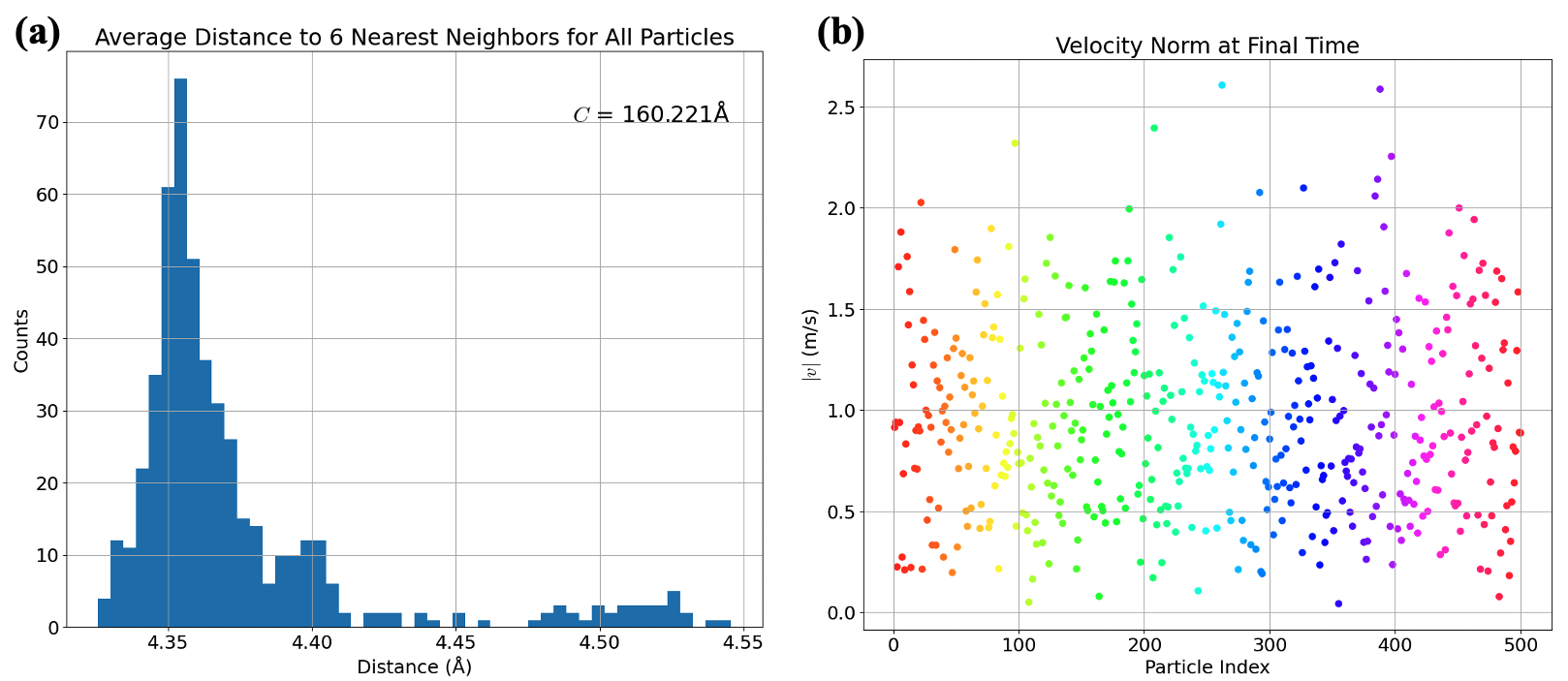}
\caption{(a) Histogram of average distance to 6 nearest neighbors for $N = 500$. This distribution is sharply peaked around $\sim4.37$\AA{}. The circumference $C$ is given for reference. (b) Velocity norm at final time step with color indicating the particle index. No particles exceed 3.5$\frac{m}{s}$, corresponding to a displacement no larger than $3.5\times10^{-15}$m for a maximum time step of $\Delta t = 10^{-15}$s. This is negligible compared to the length scale of the system.}
\label{fig:convergence}
\end{figure}

To assess convergence, we want to check that the resulting structure is lattice-like, and that particle velocity is negligible at late times. An example for $N = 500$ is shown in Figure \ref{fig:convergence}.
In Figure \ref{fig:convergence}a, we plot the average distance to each particle's six nearest neighbors. The distribution is sharply peaked near $4.37$\AA{}, and the majority of  particles attain this value within a $\pm0.05$\AA{} range. 

Figure \ref{fig:convergence}b shows the velocity of all particles at the final time step with the color indicating the particle index. No particle exceeds a velocity of $3.5$m/s, implying negligibly small displacements compared to the scale of the system over each time step. Tuning $\nu$ and increasing $N_{t}$ brings these values closer to zero, however the resulting lattice hardly changes.

\begin{figure}[ht]
\centering
\includegraphics[width=1\linewidth]{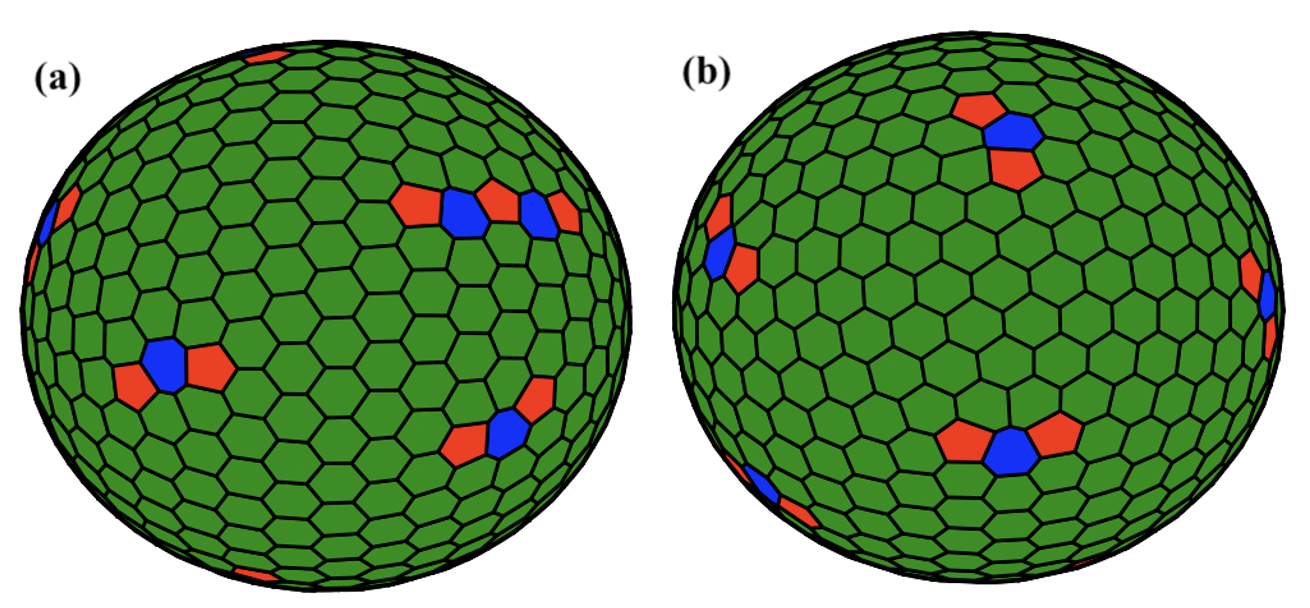}
\caption{Voronoi construction for $N = 500$ for (a) CB model and (b) Wales et al., reproduced with permission from the author.}
\label{fig:voronoi}
\end{figure}

A more quantitative metric for convergence comes from comparing our results with global minima candidates of the Thomson problem. Some of the best candidates were identified by Wales et al. in \cite{Wales_2006} and \cite{Wales_2009} using a basin hopping global optimization method. A comparison of our energy with the Wales et al. solutions for select values of $N$ is given in Table 2 in units of atomic energy.
\begin{table}[ht]
\resizebox{7.5cm}{!}{
\begin{tabular}{|c|c|c|}
\hline
\textbf{N} & \textbf{Crystal Ball} & \textbf{Wales et al.} \\ \hline
100        & 4448.4873048088       & 4448.3506343          \\ \hline
200        & 18438.93146407002     & 18438.8427175         \\ \hline
300        & 42131.54356092394     & 42131.2641372         \\ \hline
400        & 75582.9641644582      & 75582.4485122         \\ \hline
500        & 118826.43518171833    & 118825.4625421        \\ \hline
\end{tabular}}
\caption{Energy of Thomson problem solutions given in units of atomic energy.}
\end{table}
We see that our values show agreement within one unit of atomic energy of Wales et al. Additionally, our solutions exhibit similar imperfections to those of Wales et al. This can be seen by inspecting the Voronoi construction of the minimal energy configuration. An example for $N = 500$ is given in Figure \ref{fig:voronoi}.
We see from the figure that both configurations identified by the CB algorithm and Wales et al. exhibit the same pairs of imperfections: a heptagon with two adjacent hexagons. The CB model, however, exhibits one additional imperfection comprised of two heptagons and three pentagons. This could be a result of overdamping of the system, or the CB algorithm identifying a local minimum configuration. However, the close agreement in energy and heuristic similarities in the Voronoi construction suggests that we have succeeded in creating the desired spherical lattice. Our model describes average maximum energy, so this approximate agreement is sufficient for our purposes.

These results were achieved with $N_{t} = 6\times10^{4}$ steps for CB, while the values in \cite{Wales_2006} required a minimum of $10^{5}$ iterations to converge. However, this method is computationally expensive, scaling as $N^2$, though it remains feasible for small $N$. Computational efficiency could potentially be improved using a Particle-In-Cell method.

\section{Empirical Evidence of Energy Loss Proportionality to Step Size\label{appendix:C}}
\begin{figure}[ht]
    \centering
    \includegraphics[width=\linewidth]{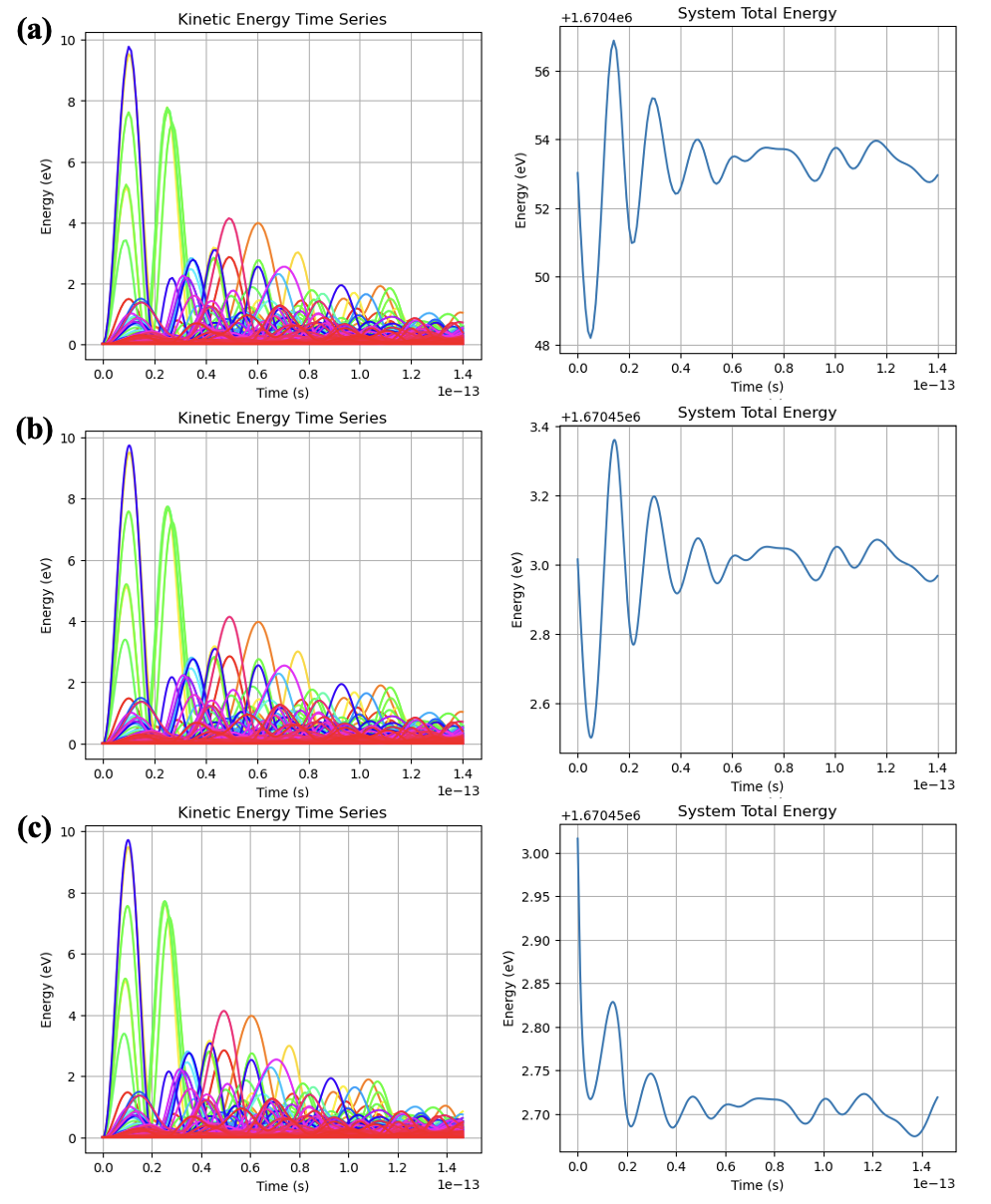}
    \caption{(a) $\Delta t = 1$ fs, $N_{t} = 140$ (b) $\Delta t = 0.1$ fs, $N_{t} = 1400$ (c) Adaptive $\Delta t$, $N_{t} = 6000$}
    \label{fig:dt_E_conservation}
\end{figure}

In Figure \ref{fig:dt_E_conservation}, we plot the energy time series for a simulation with $N = 500$ and $N_{rm} = 1$ using different time steps. In (a), (b), and (c), the times steps are $\Delta t = 1$ fs, $\Delta t = 0.1$ fs, and an adaptive time step described by equation \ref{eq:adaptiveDT} with $f = 10^{5}$, respectively. For the adaptive time step case, we also impose a maximum time step of $0.1$ fs.

We see from the total energy plots that the fluctuations, while present in all cases, decrease by an order of magnitude when changing $\Delta t$ from 1 fs to 0.1 fs. The adaptive time stepping decreases the magnitude of the fluctuations by another factor of $\sim 2$ compared to the 0.1 fs case. While all of these fluctuations are negligible compared to the total system energy, and they do not affect the computed peak energy in this case, the adaptive scheme provides a numerically stable time stepping methodology for all $N$ values.

\section{Example System Snapshots\label{appendix:B}}
\begin{figure}[H]
\centering
\includegraphics[width=1\linewidth]{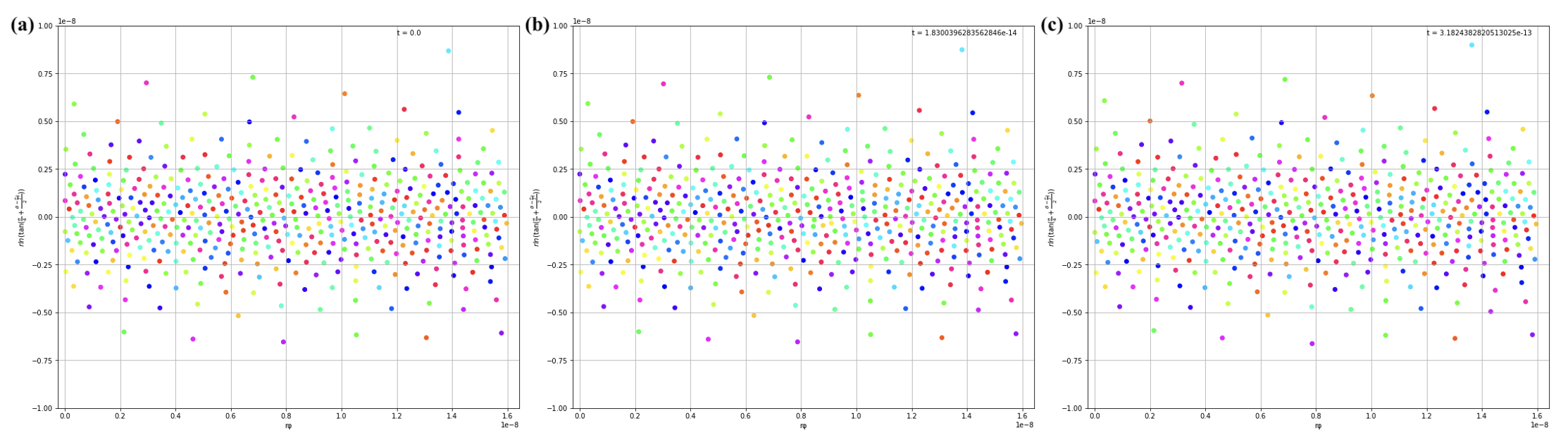}
\caption{Time snapshots of reordering of the lattice for $N = 500$, $N_{rm} = 1$. (a) Initial condition of the lattice (b) time step of closest approach (c) lattice has regained its periodic structure.}
\label{fig:figsSingle}
\end{figure}

\begin{figure}[H]
\centering
\includegraphics[width=1\linewidth]{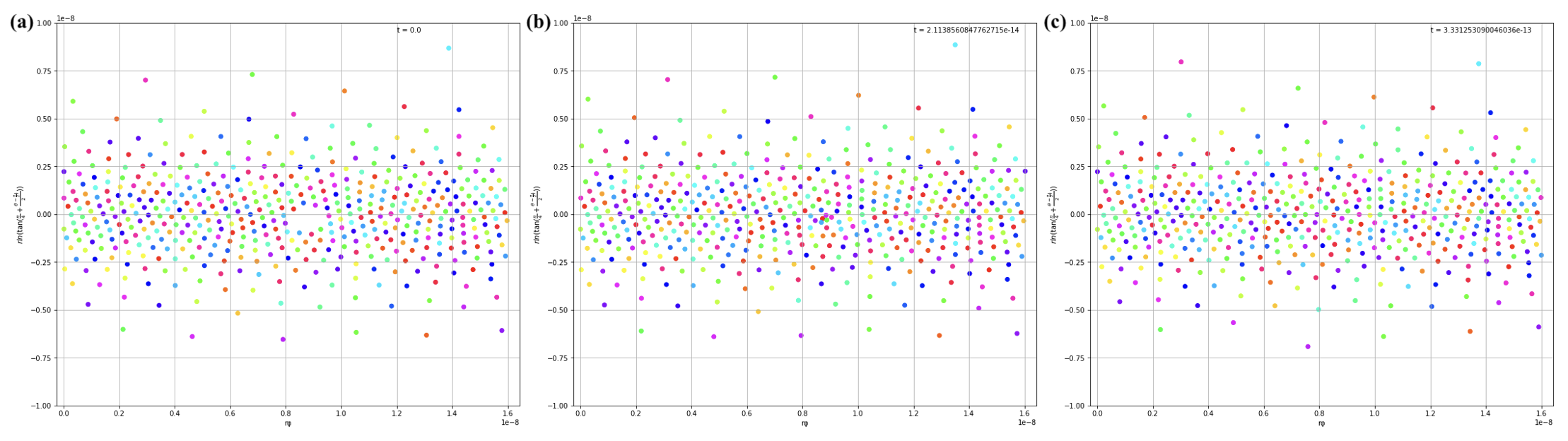}
\caption{Time snapshots of reordering of the lattice for $N = 500$, $N_{rm} = 7$. (a) Initial condition of the lattice (b) time step of closest approach (c) lattice has regained its periodic structure.}
\label{fig:figsChunk}
\end{figure}

The dynamics are most easily visualized using the Mercator projection, which is given by
\begin{equation*}
    x = r\phi
\end{equation*}
and
\begin{equation*}
    y = r\ln{\bigg[\tan{\bigg(\frac{\pi}{4}+\frac{\theta-\frac{\pi}{2}}{2}\bigg)}\bigg]}.
\end{equation*}
We shift the polar angle $\theta$ by $\frac{\pi}{2}$ in the equation above since we define this angle with respect to the positive z-axis. Snapshots for $N = 500$ with $N_{rm} = 1$ and $N_{rm} = 7$ are shown in Figures \ref{fig:figsSingle} and \ref{fig:figsChunk}, respectively.

\bibliography{apssamp}

\begin{thebibliography}{11}%
\makeatletter
\providecommand \@ifxundefined [1]{%
 \@ifx{#1\undefined}
}%
\providecommand \@ifnum [1]{%
 \ifnum #1\expandafter \@firstoftwo
 \else \expandafter \@secondoftwo
 \fi
}%
\providecommand \@ifx [1]{%
 \ifx #1\expandafter \@firstoftwo
 \else \expandafter \@secondoftwo
 \fi
}%
\providecommand \natexlab [1]{#1}%
\providecommand \enquote  [1]{``#1''}%
\providecommand \bibnamefont  [1]{#1}%
\providecommand \bibfnamefont [1]{#1}%
\providecommand \citenamefont [1]{#1}%
\providecommand \href@noop [0]{\@secondoftwo}%
\providecommand \href [0]{\begingroup \@sanitize@url \@href}%
\providecommand \@href[1]{\@@startlink{#1}\@@href}%
\providecommand \@@href[1]{\endgroup#1\@@endlink}%
\providecommand \@sanitize@url [0]{\catcode `\\12\catcode `\$12\catcode `\&12\catcode `\#12\catcode `\^12\catcode `\_12\catcode `\%12\relax}%
\providecommand \@@startlink[1]{}%
\providecommand \@@endlink[0]{}%
\providecommand \url  [0]{\begingroup\@sanitize@url \@url }%
\providecommand \@url [1]{\endgroup\@href {#1}{\urlprefix }}%
\providecommand \urlprefix  [0]{URL }%
\providecommand \Eprint [0]{\href }%
\providecommand \doibase [0]{https://doi.org/}%
\providecommand \selectlanguage [0]{\@gobble}%
\providecommand \bibinfo  [0]{\@secondoftwo}%
\providecommand \bibfield  [0]{\@secondoftwo}%
\providecommand \translation [1]{[#1]}%
\providecommand \BibitemOpen [0]{}%
\providecommand \bibitemStop [0]{}%
\providecommand \bibitemNoStop [0]{.\EOS\space}%
\providecommand \EOS [0]{\spacefactor3000\relax}%
\providecommand \BibitemShut  [1]{\csname bibitem#1\endcsname}%
\let\auto@bib@innerbib\@empty
\bibitem [{\citenamefont {Hoang}\ and\ \citenamefont {Giang}(2019)}]{Hoang_2019}%
  \BibitemOpen
  \bibfield  {author} {\bibinfo {author} {\bibfnamefont {V.~V.}\ \bibnamefont {Hoang}}\ and\ \bibinfo {author} {\bibfnamefont {N.~H.}\ \bibnamefont {Giang}},\ }\bibfield  {title} {\bibinfo {title} {Compression-induced square-triangle solid-solid phase transition in 2d simple monatomic system},\ }\href {https://doi.org/https://doi.org/10.1016/j.physe.2019.05.001} {\bibfield  {journal} {\bibinfo  {journal} {Physica E: Low-dimensional Systems and Nanostructures}\ }\textbf {\bibinfo {volume} {113}},\ \bibinfo {pages} {35} (\bibinfo {year} {2019})}\BibitemShut {NoStop}%
\bibitem [{\citenamefont {Wales}\ and\ \citenamefont {Ulker}(2006)}]{Wales_2006}%
  \BibitemOpen
  \bibfield  {author} {\bibinfo {author} {\bibfnamefont {D.~J.}\ \bibnamefont {Wales}}\ and\ \bibinfo {author} {\bibfnamefont {S.}~\bibnamefont {Ulker}},\ }\bibfield  {title} {\bibinfo {title} {Structure and dynamics of spherical crystals characterized for the thomson problem},\ }\href {https://doi.org/10.1103/PhysRevB.74.212101} {\bibfield  {journal} {\bibinfo  {journal} {Phys. Rev. B}\ }\textbf {\bibinfo {volume} {74}},\ \bibinfo {pages} {212101} (\bibinfo {year} {2006})}\BibitemShut {NoStop}%
\bibitem [{\citenamefont {Wales}\ \emph {et~al.}(2009)\citenamefont {Wales}, \citenamefont {McKay},\ and\ \citenamefont {Altschuler}}]{Wales_2009}%
  \BibitemOpen
  \bibfield  {author} {\bibinfo {author} {\bibfnamefont {D.~J.}\ \bibnamefont {Wales}}, \bibinfo {author} {\bibfnamefont {H.}~\bibnamefont {McKay}},\ and\ \bibinfo {author} {\bibfnamefont {E.~L.}\ \bibnamefont {Altschuler}},\ }\bibfield  {title} {\bibinfo {title} {Defect motifs for spherical topologies},\ }\href {https://doi.org/10.1103/PhysRevB.79.224115} {\bibfield  {journal} {\bibinfo  {journal} {Phys. Rev. B}\ }\textbf {\bibinfo {volume} {79}},\ \bibinfo {pages} {224115} (\bibinfo {year} {2009})}\BibitemShut {NoStop}%
\bibitem [{\citenamefont {Moscato}\ \emph {et~al.}(2023)\citenamefont {Moscato}, \citenamefont {Haque},\ and\ \citenamefont {Moscato}}]{Moscato_2023}%
  \BibitemOpen
  \bibfield  {author} {\bibinfo {author} {\bibfnamefont {P.}~\bibnamefont {Moscato}}, \bibinfo {author} {\bibfnamefont {M.}~\bibnamefont {Haque}},\ and\ \bibinfo {author} {\bibfnamefont {A.}~\bibnamefont {Moscato}},\ }\bibfield  {title} {\bibinfo {title} {Continued fractions and the thomson problem},\ }\href {https://doi.org/10.1038/s41598-023-33744-5} {\bibfield  {journal} {\bibinfo  {journal} {Scientific Reports}\ }\textbf {\bibinfo {volume} {13}} (\bibinfo {year} {2023})}\BibitemShut {NoStop}%
\bibitem [{\citenamefont {Lai}\ \emph {et~al.}(2024)\citenamefont {Lai}, \citenamefont {Hao}, \citenamefont {Xiao},\ and\ \citenamefont {Fu}}]{Lai_2024}%
  \BibitemOpen
  \bibfield  {author} {\bibinfo {author} {\bibfnamefont {X.}~\bibnamefont {Lai}}, \bibinfo {author} {\bibfnamefont {J.-K.}\ \bibnamefont {Hao}}, \bibinfo {author} {\bibfnamefont {R.}~\bibnamefont {Xiao}},\ and\ \bibinfo {author} {\bibfnamefont {Z.-H.}\ \bibnamefont {Fu}},\ }\bibfield  {title} {\bibinfo {title} {Global optimization and structural analysis of coulomb and logarithmic potentials on the unit sphere using a population-based heuristic approach},\ }\href {https://doi.org/https://doi.org/10.1016/j.eswa.2023.121704} {\bibfield  {journal} {\bibinfo  {journal} {Expert Systems with Applications}\ }\textbf {\bibinfo {volume} {238}},\ \bibinfo {pages} {121704} (\bibinfo {year} {2024})}\BibitemShut {NoStop}%
\bibitem [{\citenamefont {Balantekin}\ and\ \citenamefont {Takigawa}(1998)}]{Balantekin_1998}%
  \BibitemOpen
  \bibfield  {author} {\bibinfo {author} {\bibfnamefont {A.~B.}\ \bibnamefont {Balantekin}}\ and\ \bibinfo {author} {\bibfnamefont {N.}~\bibnamefont {Takigawa}},\ }\bibfield  {title} {\bibinfo {title} {Quantum tunneling in nuclear fusion},\ }\href {https://doi.org/10.1103/revmodphys.70.77} {\bibfield  {journal} {\bibinfo  {journal} {Reviews of Modern Physics}\ }\textbf {\bibinfo {volume} {70}},\ \bibinfo {pages} {77–100} (\bibinfo {year} {1998})}\BibitemShut {NoStop}%
\bibitem [{Note1()}]{Note1}%
  \BibitemOpen
  \bibinfo {note} {The Thomson problem originally addresses electrons on a positive equipotential. Here, we consider a system of positive, non-relativistic point charges.}\BibitemShut {Stop}%
\bibitem [{\citenamefont {Tentori}\ and\ \citenamefont {Belloni}(2023)}]{Tentori_2023}%
  \BibitemOpen
  \bibfield  {author} {\bibinfo {author} {\bibfnamefont {A.}~\bibnamefont {Tentori}}\ and\ \bibinfo {author} {\bibfnamefont {F.}~\bibnamefont {Belloni}},\ }\bibfield  {title} {\bibinfo {title} {Revisiting p-11b fusion cross section and reactivity, and their analytic approximations},\ }\href {https://doi.org/10.1088/1741-4326/acda4b} {\bibfield  {journal} {\bibinfo  {journal} {Nuclear Fusion}\ }\textbf {\bibinfo {volume} {63}},\ \bibinfo {pages} {086001} (\bibinfo {year} {2023})}\BibitemShut {NoStop}%
\bibitem [{\citenamefont {Kaneti}\ \emph {et~al.}(2022)\citenamefont {Kaneti} \emph {et~al.}}]{Kaneti_2022}%
  \BibitemOpen
  \bibfield  {author} {\bibinfo {author} {\bibfnamefont {Y.~V.}\ \bibnamefont {Kaneti}} \emph {et~al.},\ }\bibfield  {title} {\bibinfo {title} {Borophene: Two-dimensional boron monolayer: Synthesis, properties, and potential applications},\ }\href {https://doi.org/10.1021/acs.chemrev.1c00233} {\bibfield  {journal} {\bibinfo  {journal} {Chemical Reviews}\ }\textbf {\bibinfo {volume} {122}},\ \bibinfo {pages} {1000} (\bibinfo {year} {2022})},\ \bibinfo {note} {pMID: 34730341},\ \Eprint {https://arxiv.org/abs/https://doi.org/10.1021/acs.chemrev.1c00233} {https://doi.org/10.1021/acs.chemrev.1c00233} \BibitemShut {NoStop}%
\bibitem [{\citenamefont {Luo}\ \emph {et~al.}(2017)\citenamefont {Luo}, \citenamefont {Fan},\ and\ \citenamefont {An}}]{Zhifen_2017}%
  \BibitemOpen
  \bibfield  {author} {\bibinfo {author} {\bibfnamefont {Z.}~\bibnamefont {Luo}}, \bibinfo {author} {\bibfnamefont {X.}~\bibnamefont {Fan}},\ and\ \bibinfo {author} {\bibfnamefont {Y.}~\bibnamefont {An}},\ }\bibfield  {title} {\bibinfo {title} {First-principles study on the stability and stm image of borophene},\ }\href {https://doi.org/10.1186/s11671-017-2282-7} {\bibfield  {journal} {\bibinfo  {journal} {Nanoscale Research Letters}\ }\textbf {\bibinfo {volume} {12}} (\bibinfo {year} {2017})}\BibitemShut {NoStop}%
\bibitem [{\citenamefont {Burke}(2011)}]{Burke_2011}%
  \BibitemOpen
  \bibfield  {author} {\bibinfo {author} {\bibfnamefont {P.~G.}\ \bibnamefont {Burke}},\ }\href@noop {} {\emph {\bibinfo {title} {R-matrix theory of atomic collisions application to atomic, molecular and Optical Processes}}}\ (\bibinfo  {publisher} {Springer Berlin Heidelberg},\ \bibinfo {year} {2011})\BibitemShut {NoStop}%
\end{thebibliography}%
\end{document}